# Copper-catalyzed efficient synthesis of 5-arylindazolo[3,2-*b*]quinazolin-7(5*H*)-ones from 2-nitrobenzaldehydes

Zahra ArastehFard [a], Karim Akbari Dilmaghani [a], Mehdi Soheilizad [b] and Mohammad Mahdavi [c]*

[a] *Department of Chemistry, Faculty of Sciences, Urmia University, Urmia, Iran*
[b] *School of Chemistry, College of Science, University of Tehran, Tehran, Iran*
[c] *Endocrinology and Metabolism Research Center, Endocrinology and Metabolism Clinical Sciences Institute, Tehran University of Medical Sciences, Tehran, Iran*

**Abstract**

A novel and practical copper-catalyzed approach is developed for the preparation of 5-arylindazolo[3,2-*b*]quinazolin-7(5*H*)-ones. The 2-amino-*N*′-arylbenzohydrazide, which easily prepared by reaction of isatoic anhydride with arylhydrazine, through a condensation/intramolecular cyclization reacted by 2-nitrobenzaldehydes in the present of CuI to afford corresponding 5-arylindazolo[3,2-*b*]quinazolin-7(5*H*)-ones in good yields.

***Keywords:*** *Indazolo[3,2-b]quinazolinones; copper-catalyzed; 2-amino-N'-arylbenzohydrazide; 2-nitrobenzaldehyde; intramolecular cyclization.*

## 1. Introduction

Nitrogen-containing heterocyclic compounds and their analogous are pharmaceutically attractive scaffolds and widely exist in naturally occurring and synthetic biologically active molecules.[1] Among them, fused polycyclic *N*-containing heteroaromatics have received much synthetic attention because of their wide range of biological activities and their high therapeutic values.[2] For instance, *N*-fused polycycles having the quinazolinone scaffold have been shown to possess a broad range of biological activities, including anticancer,[3] anti-microbial,[4] anti-inflammatory,[5] anticonvulsant,[6] anti-ulcer,[7] anti-bacterial,[8] antidiabetic,[9] and anti-virus.[10] They are also an important class of alkaloids because they are widely found in the structure of a number of natural products such as luotonin A,[11] circumdatins,[12] rutaecarpine,[13] deoxyvasicinone,[14] and tryptanthrin.[15] Another instance of *N*-containing biologically active heterocycles is indazole-based derivatives which have been reported to possess antidepressant,[16] anti-inflammatory,[16] HIV protease inhibition,[17] antitumor,[18] antimicrobial,[19] and contraceptive activities.[20] The indazole core is an important pharmacophore in medicinal chemistry and has been recognized as a privileged structure in heterocyclic chemistry.

Owing to this broad range of properties of *N*-containing quinazolinone and indazole heterocycles, it is reasonable to expect that fused quinazolinone-indazole derivatives, such as indazolo[3,2-*b*]quinazolinones have significant biological activity. To date, a few synthetic routes have been reported for the preparation of indazolo[3,2-*b*]quinazolinones including: one-pot cascade reaction of isatoic anhydride, hydrazines and 2-iodo benzaldehyde catalyzed by palladium,[21] Cu-catalyzed domino Ullmann-type coupling reaction of 2-amino-*N*′-arylbenzohydrazide and 2-halobenzaldehydes,[22] intramolecular C-N bond formation reaction of 3-amino-2-(2- bromophenyl) dihydroquinazolinones by CuCl/L-proline catalytic system,[23] and Pd-catalyzed cascade reaction of 2-amino-*N*′-arylbenzohydrazides with triethyl orthobenzoates.[24] So, we wish to developed these procedures via intramolecular cyclization of 2-amino-*N*′-arylbenzohydrazide with 2-nitrobenzaldehydes catalyzed by copper(I) iodide (Scheme 1).

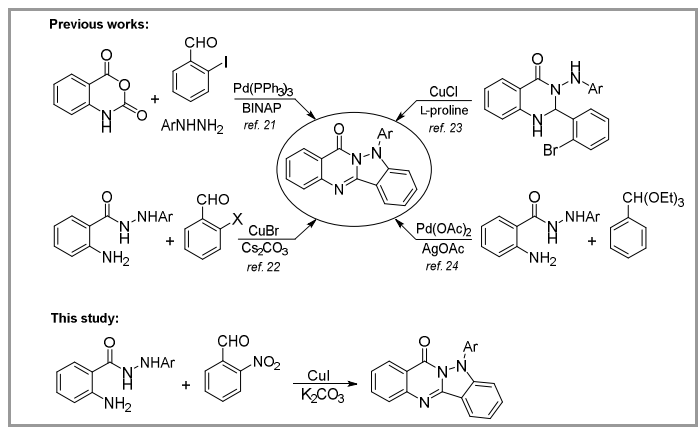

**Scheme 1.** Synthesis of 5-arylindazolo[3,2-*b*]quinazolin-7(5*H*)-ones

## 2. Results and discussion

In continuation of our efforts for the efficient synthesis of biologically active target molecules,[25] herein we would like to introduce a new approach for the preparation of 5-arylindazolo[3,2-*b*]quinazolin-7(5*H*)-ones. Thus, initially 2-amino-*N*′-arylbenzohydrazide **3** were easily prepared by the reaction of isatoic anhydride **1** and arylhydrazine **2** in aqueous media.[25g] Next, heating an equimolar mixture of 2-amino-*N*′-arylbenzohydrazide **3** and 2-nitrobenzaldehydes **4**, in the present of CuI and $K_2CO_3$ in DMSO at 80 °C for 8 hours, through a condensation and intramolecular cyclization by removal of nitro group,[26] leads to formation of corresponding 5-arylindazolo[3,2-*b*]quinazolin-7(5*H*)-ones **5a-l** in good yields (Scheme 2).

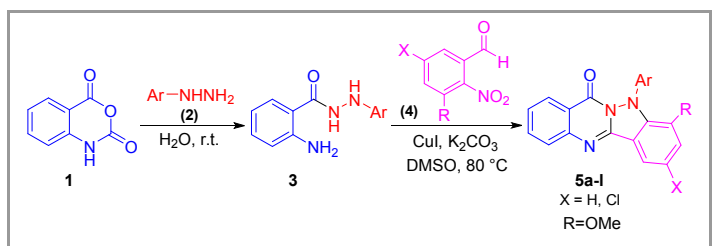

**Scheme 2.** Copper-catalyzed preparation of 5-arylindazolo[3,2-*b*]quinazolin-7(5*H*)-ones

In order to optimization of reaction conditions, preparation of 5-phenylindazolo[3,2-*b*]quinazolin-7(5*H*)-one (**5a**) was investigated as model reaction. At first, we observed that heating a mixture of 2-amino-*N*′-phenylbenzohydrazide and 2-nitrobenzaldehyde in the present of CuBr and $Cs_2CO_3$ in DMF as solvent, at 80 °C for 8 hours leads to desire product **5a** in 58% yield (Table 1, entry 1). Next, to evaluate of reaction medium, the effect of the several solvents such as DMSO, xylene and THF was investigated (Table 1, entries 2–4), that the best result was obtained in DMSO (65%, Table 1, entry 2). Then, to choice of the best base, the model reaction was examined by various

―――――
* Corresponding author. Tel.: +98-21-66954708; fax: +98-21-66461178; e-mail: Mahdavi_chem@yahoo.com

inorganic and organic bases such as K₂CO₃, K₃PO₄, DBU and NEt₃ (Table 1, entries 5–8). As shown in Table 1, among various bases examined, K₂CO₃ turned out to be the best choice, while others were less effective. Next, to find the best copper source, we observed that the use of CuI as Cu-source significantly increases the yield of **5a** (Table 1, entry 9), while the other Cu salts like CuCl, CuO, Cu₂O, and Cu(OAc)₂ could not enhance the yield of desire product **5a** (Table 1, entries 10−13). Finally, to examine the effect of temperature, the model reaction in the present of CuI and K₂CO₃ in DMSO was investigated in several different temperatures (Table 1, entries 14−16). As it could be seen in Table 1, the best yield of the product is obtained at 80 °C (85%, Table 1, entry 9).

**Table 1**. Optimization of conditions for the synthesis of **5a**[a]

| Entry | Cu source | Base | Solvent | Temp. (°C) | Yield (%)[b] |
|---|---|---|---|---|---|
| 1 | CuBr | CsCO₃ | DMF | 80 | 58 |
| 2 | CuBr | CsCO₃ | DMSO | 80 | 65 |
| 3 | CuBr | CsCO₃ | Xylene | 80 | NR[c] |
| 4 | CuBr | CsCO₃ | THF | 80 | 32 |
| 5 | CuBr | K₂CO₃ | DMSO | 80 | 68 |
| 6 | CuBr | K₃PO₄ | DMSO | 80 | 48 |
| 7 | CuBr | DBU | DMSO | 80 | 42 |
| 8 | CuBr | NEt₃ | DMSO | 80 | 37 |
| **9** | **CuI** | **K₂CO₃** | **DMSO** | **80** | **85** |
| 10 | CuCl | K₂CO₃ | DMSO | 80 | 46 |
| 11 | CuO | K₂CO₃ | DMSO | 80 | 18 |
| 12 | Cu₂O | K₂CO₃ | DMSO | 80 | 38 |
| 13 | Cu(OAc)₂ | K₂CO₃ | DMSO | 80 | 22 |
| 14 | CuI | K₂CO₃ | DMSO | 120 | 78 |
| 15 | CuI | K₂CO₃ | DMSO | 60 | 74 |
| 16 | CuI | K₂CO₃ | DMSO | 25 | 15 |
| 17 | – | K₂CO₃ | DMSO | 80 | NR |

[a] Reaction condition: **3a** (2 mmol), **4a** (2.1 mmol), Cu-source (0.2 mmol), Base (2 mmol), Solvent (4 mL) at reflux for 8 h. [b] Isolated yield. [c] NR= no reaction

To investigate the scope of this reaction, the reaction between 2-amino-*N'*-arylbenzohydrazide **3** and various 2-nitrobenzaldehydes **4**, carrying both electron-donating and electron-withdrawing substituent, were explored under the optimized conditions to afford the 5-arylindazolo[3,2-*b*]quinazolin-7(5*H*)-ones **5a–l** in 69–88% yields (Table 2).

**Table 2.** Substrate scope of Cu-catalyzed synthesis of 5-arylindazolo[3,2-*b*]quinazolin-7(5*H*)-ones **5a–l**

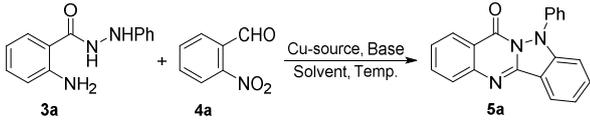

A possible reaction pathway for the Cu-catalyzed formation of 5-arylindazolo[3,2-*b*]quinazolin-7(5*H*)-ones **5** is suggested in Scheme 3. The initial condensation of the 2-amino-*N'*-arylbenzohydrazide **3** with 2-nitrobenzaldehyde **4**, followed by aerobic oxidation generate the compound **6** which undergo coordination with Cu generates intermediate **7**. Then, nucleophilic attack of iodide ion to this intermediate followed by intramolecular cyclization and libration of KNO₂, leads to the formation 5-arylindazolo[3,2-*b*]quinazolin-7(5*H*)-ones **5** (Scheme 3).

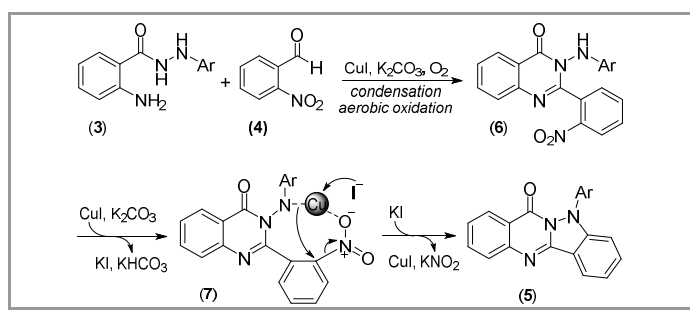

**Scheme 3.** Proposed mechanism for the preparation of 5-arylindazolo[3,2-*b*]quinazolin-7(5*H*)-ones

### 3. Conclusion

In conclusion, we develop a new and efficient copper-catalyzed approach for the synthesis of 5-arylindazolo[3,2-*b*]quinazolin-7(5*H*)-ones through a condensation/intramolecular cyclization reaction of 2-amino-*N'*-arylbenzohydrazide with 2-

nitrobenzaldehydes in the present of CuI. The simplicity of the starting materials, ligand-free metal catalyzed and good yields of the products are the main advantages of this method.

## 4. Experimental

### 4.1. Material and methods

All chemicals were purchased from Merck and Fluka companies. All yields refer to isolated products. Melting points were determined in a capillary tube and are not corrected. The progress of reaction was followed with TLC using silica gel SILG/UV 254 and 365 plates. IR spectra of the compounds were obtained on Nicolet FT-IR Magna 550 spectrographs (KBr disks) (Nicolet, Madison, WI, USA). $^1$H and $^{13}$C NMR spectra were recorded on a Brucker, Rheinstetten, Germany (at 500 and 125 MHz) NMR spectrometer using tetramethylsilane (TMS) as internal standard. Products **5e**, **5f**, **5h**, **5i**, **5j**, **5k** and **5l** are novel compounds. All products are characterized by melting point, IR, $^1$H and $^{13}$C NMR spectroscopic data.

### 4.2. General procedure for the preparation of 2-amino-N′-arylbenzohydrazide (3)

A mixture of isatoic anhydride (2.0 mmol) and arylhydrazine (2.0 mmol) in $H_2O$ (4.0 mL) was stirred for 4 hours at room temperature. After complete the reaction as indicated by TLC, the resulting precipitate was filtered, washed with cold water, dried, and recrystallized from ethanol to afford the desired compounds **3**.

### 4.3. General procedure for the preparation of 5-arylindazolo[3,2-b]quinazolin-7(5H)-ones (5a-i)

A mixture of 2-amino-N′-arylbenzohydrazide **3** (2.0 mmol), 2-nitrobenzaldehyde **4** (2.1 mmol), CuI (0.2 mmol), and $K_2CO_3$ (2.0 mmol) in dry DMSO (4.0 mL) was heated in a sealed vessel under air for 8 hours at 80 °C. After reaction completion (TLC), the reaction mixture was cooled, quenched with water (20 mL), and extracted with EtOAc (3 × 20 mL). The extract was washed with 20% NaCl solution (W/V), dried ($Na_2SO_4$) and concentrated under reduced pressure. The residue was purified by column chromatography using petroleum ether/ethyl acetate (4:1) as eluent to afford the pure product **5a–l**.

#### 4.3.1. 5-phenylindazolo[3,2-b]quinazolin-7(5H)-one (5a).
Yield 85% (264 mg); White solid; M.p = 246–248 °C (Lit. 230–231 °C)[24]; IR (KBr, cm$^{-1}$): 3052, 1672, 1623, 1469, 1054; $^1$H NMR (CDCl$_3$, 500 MHz): δ = 7.19 (d, J = 8.0 Hz, 1H), 7.36 (d, J = 7.5 Hz, 2H), 7.4 (t, J = 7.5 Hz, 2H), 7.44 (m, 3H), 7.59 (t, J = 7.5 Hz, 1H), 7.80 (t, J = 7.5 Hz, 1H), 7.9 (d, J = 8.0 Hz, 1H), 8.27 (d, J = 8.0 Hz, 1H), 8.32 (d, J = 7.5 Hz, 1H); $^{13}$C NMR (CDCl$_3$, 125 MHz): δ = 112.3, 118.8, 119.7, 123.2, 124.2, 124.5, 125.3, 126.6, 126.9, 128.5, 129.4, 133.3, 133.9, 141.8, 148.1, 148.6, 149.0, 156.3; Anal. Calcd for $C_{20}H_{13}N_3O$: C, 77.16; H, 4.21; N, 13.50. Found: C, 76.12; H, 4.17; N, 13.12.

#### 4.3.2. 5-(p-tolyl)indazolo[3,2-b]quinazolin-7(5H)-one (5b).
Yield 87% (283 mg); White solid; M.p = 196–198 °C (Lit. 184–185 °C)[24]; IR (KBr, cm$^{-1}$): 3030, 1688, 1619, 1464, 1057; $^1$H NMR (DMSO-$d_6$, 500 MHz): δ = 2.36 (s, 3H), 7.27 (d, J = 7.5 Hz, 1H), 7.29 (d, J = 8.5 Hz, 2H), 7.33 (d, J = 8.0 Hz, 2H), 7.48 (t, J = 7.5 Hz, 1H), 7.52 (m, 2H), 7.73 (t, J = 7.5 Hz, 1H), 7.87 (d, J = 6.5 Hz, 1H), 8.16 (d, J = 7.5 Hz,1H), 8.23 (d, J = 7.5 Hz,1H); $^{13}$C NMR (DMSO-$d_6$, 125 MHz): δ = 20.5, 112.5, 118.2, 119.4, 122.9, 123.7, 123.9, 124.4, 124.6, 125.1, 125.3, 125.8, 125.9, 129.8, 133.7, 133.9, 137.4, 139.2, 148.2, 148.6, 155.2; Anal. Calcd for $C_{21}H_{15}N_3O$: C, 77.52; H, 4.65; N, 12.91. Found: C, 77.47; H, 4.52; N, 12.82.

#### 4.3.3. 5-(4-methoxyphenyl)indazolo[3,2-b]quinazolin-7(5H)-one (5c).
Yield 88% (301 mg); White solid; M.p = 216–218 °C (Lit. 209–210 °C)[23]; IR (KBr, cm$^{-1}$): 3053, 1674, 1624, 1468, 1055; $^1$H NMR (CDCl$_3$, 500 MHz): δ = 3.84 (s, 3H), 6.96 (d, J = 8.0 Hz, 2H), 7.11 (d, J = 8.0 Hz, 1H), 7.3 (d, J = 8.0 Hz, 2H), 7.39 (t, J = 7.5 Hz, 1H), 7.44 (t, J = 7.5 Hz, 1H), 7.59 (t, J = 7.5 Hz, 1H), 7.79 (t, J = 7.5 Hz, 1H), 7.89 (d, J = 7.5 Hz, 1H), 8.27 (d, J = 7.5 Hz,1H), 8.32 (d, J = 7.5 Hz, 1H); $^{13}$C NMR (CDCl$_3$, 125 MHz): δ = 55.5, 112.4, 114.5, 118.6, 119.8, 123.1, 124.0, 125.2, 126.6, 127.0, 133.2, 133.8, 134.4, 148.0, 148.6, 149.5, 156.4, 159.6; Anal. Calcd for $C_{21}H_{15}N_3O_2$: C, 73.89; H, 4.43; N, 12.31. Found: C, 73.68; H, 4.22; N, 12.13.

#### 4.3.4. 2-chloro-5-phenylindazolo[3,2-b]quinazolin-7(5H)-one (5d). 
Yield 83% (287 mg); White solid; M.p = 241–243 °C (Lit. 272–273 °C)[22]; IR (KBr, cm$^{-1}$): 3184, 1679, 1627, 1465, 1053; $^1$H NMR (CDCl$_3$, 500 MHz): δ = 7.12 (d, J = 8.5 Hz, 1H), 7.35 (d, J = 7.5 Hz, 2H), 7.42 (t, J = 7.0 Hz, 1H), 7.47 (m, 3H), 7.55 (d, J = 8.5 Hz, 1H), 7.82 (t, J = 7.5 Hz, 1H), 7.9 (d, J = 8.0 Hz, 1H), 8.26 (s, J = 8.0 Hz, 1H), 8.32 (d, J = 7.5 Hz, 1H); $^{13}$C NMR (CDCl$_3$, 125 MHz): δ = 113.6, 119.9, 120.2, 122.8, 124.5, 125.7, 126.7, 127.1, 128.8, 129.6, 130.0, 133.6, 134.1, 141.4, 147.1, 147.7, 148.4, 148.9, 155.2, 156.2; Anal. Calcd for $C_{20}H_{12}ClN_3O$: C, 69.47; H, 3.50; N, 12.15. Found: C, 69.25; H, 3.21; N, 11.94.

#### 4.3.5. 2-chloro-5-(p-tolyl)indazolo[3,2-b]quinazolin-7(5H)-one (5e).
Yield 86% (309 mg); White solid; M.p = 226–228 °C; IR (KBr, cm$^{-1}$): 3020, 1682, 1626, 1466, 1054; $^1$H NMR (CDCl$_3$, 500 MHz): δ = 2.40 (s, 3H), 7.09 (d, J = 8.5 Hz, 1H), 7.23 (d, J = 7.5 Hz, 2H), 7.27 (d, J = 9.0 Hz, 2H), 7.46 (t, J = 7.5 Hz, 1H), 7.54 (d, J = 9.0 Hz, 1H), 7.81 (t, J = 7.5 Hz, 1H), 7.89 (d, J = 8.0 Hz, 1H), 8.25 (s, 1H), 8.32 (d, J = 8.0 Hz, 1H); $^{13}$C NMR (CDCl$_3$, 125 MHz): δ = 21.2, 113.6, 119.9, 122.7, 124.7, 125.6, 126.7, 126.7, 127.0, 127.1, 130.2, 133.5, 134.0, 138.8, 139.0, 147.6, 148.6, 156.3; Anal. Calcd for $C_{21}H_{14}ClN_3O$: C, 70.10; H, 3.92; N, 11.68. Found: C, 69.91; H, 3.58; N, 11.35.

#### 4.3.6. 2-chloro-5-(4-methoxyphenyl)indazolo[3,2-b]quinazolin-7(5H)-one (5f).
Yield 85% (319 mg); White solid; M.p = 238–240 °C; IR (KBr, cm$^{-1}$): 3046, 1681, 1627, 1466, 1060; $^1$H NMR (CDCl$_3$, 500 MHz): δ = 3.84 (s, 3H), 6.97 (d, J = 8.0 Hz, 2H), 7.05 (d, J = 9.0 Hz, 1H), 7.29 (d, J = 8.5 Hz, 2H), 7.46 (t, J = 7.5 Hz, 1H), 7.55 (d, J = 8.5 Hz, 1H), 7.81 (t, J = 7.5 Hz, 1H), 7.89 (d, J = 8.0 Hz, 1H), 8.25 (s, 1H), 8.32 (d, J = 7.5 Hz, 1H); $^{13}$C NMR (CDCl$_3$, 125 MHz): δ = 55.5, 113.7, 114.7, 117.5, 121.3, 125.0, 125.3, 125.6, 126.8, 126.9, 128.7, 129.8, 133.6, 134.0, 147.7, 156.7, 157.0, 157.3; Anal. Calcd for $C_{21}H_{14}ClN_3O_2$: C, 67.12; H, 3.76; N, 11.18. Found: C, 66.87; H, 3.48; N, 10.93.

#### 4.3.7. 5-(4-fluorophenyl)indazolo[3,2-b]quinazolin-7(5H)-one (5g).
Yield 83% (273 mg); White solid; M.p = 213–215 °C (Lit. 197–198 °C)[23]; IR (KBr, cm$^{-1}$): 3076, 1674, 1626, 1467, 1059; $^1$H NMR (CDCl$_3$, 500 MHz): δ = 7.15 (m, 3H), 7.37 (d, J = 8.0 Hz, 2H), 7.43 (t, J = 7.0 Hz, 1H), 7.46 (t, J = 7.0 Hz, 1H), 7.63 (t, J = 7.5 Hz, 1H), 7.82 (t, J = 7.5 Hz, 1H), 7.91 (d, J = 8.0 Hz, 1H), 8.29 (d, J = 8.0 Hz, 1H), 8.32 (d, J = 8.0 Hz, 1H); $^{13}$C NMR (CDCl$_3$, 125 MHz): δ = 112.3, 116.4, 116.5, 118.9, 123.5, 124.4, 125.5, 126.6, 127.0, 127.1, 130.3, 133.4, 134.0, 143.7, 148.6,

153.6, 155.3; Anal. Calcd for C$_{20}$H$_{12}$FN$_3$O: C, 72.94; H, 3.67; N, 12.76. Found: C, 72.52; H, 3.44; N, 12.35.

4.3.8. *2-chloro-5-(4-fluorophenyl)indazolo[3,2-b]quinazolin-7(5H)-one* (**5h**). Yield 77% (281 mg); White solid; M.p = 222–224 °C; IR (KBr, cm$^{-1}$): 3056, 1683, 1625, 1465, 1057; $^1$H NMR (CDCl$_3$, 500 MHz): δ = 7.06 (d, *J* = 8.5 Hz, 1H), 7.17 (d, *J* = 7.5 Hz, 2H), 7.35 (d, *J* = 8.0 Hz, 2H), 7.48 (t, *J* = 7.5 Hz, 1H), 7.57 (d, *J* = 8.5 Hz, 1H), 7.82 (t, *J* = 7.5 Hz, 1H), 7.89 (d, *J* = 8.0 Hz, 1H), 8.26 (s, 1H), 8.31 (d, *J* = 8.0 Hz, 1H); $^{13}$C NMR (CDCl$_3$, 125 MHz): δ = 113.6, 116.5, 116.7, 125.3, 125.4, 125.8, 125.8, 127.0, 127.1, 127.2, 130.9, 133.7, 134.3, 134.3, 136.0, 143.9, 153.1, 155.3; Anal. Calcd for C$_{20}$H$_{11}$ClFN$_3$O: C, 66.04; H, 3.05; N, 11.55. Found: C, 65.72; H, 2.87; N, 11.23.

4.3.9. *5-(4-bromophenyl)-2-chloroindazolo[3,2-b]quinazolin-7(5H)-one* (**5i**). Yield 80% (340 mg); White solid; M.p = 244–246 °C; IR (KBr, cm$^{-1}$): 3061, 1678, 1629, 1465, 1052; $^1$H NMR (DMSO-*d*$_6$, 500 MHz): δ = 7.39 (d, *J* = 9.0 Hz, 1H), 7.49 (d, *J* = 8.0 Hz, 2H), 7.55 (t, *J* = 7.5 Hz, 1H), 7.68 (d, *J* = 8.0 Hz, 2H), 7.78 (d, *J* = 8.5 Hz, 1H), 7.89 (t, *J* = 7.5 Hz, 1H), 7.93 (d, *J* = 8.0 Hz, 1H), 8.17 (d, *J* = 8.0 Hz, 1H), 8.27 (s, 1H); $^{13}$C NMR (CDCl$_3$, 125 MHz): δ = 114.0, 120.8, 122.2, 124.3, 124.6, 124.7, 125.6, 126.0, 126.0, 126.4, 128.9, 132.2, 133.7, 140.4, 146.4, 147.9, 155.5; Anal. Calcd for C$_{20}$H$_{11}$BrClN$_3$O: C, 56.56; H, 2.61; N, 9.89. Found: C, 56.31; H, 2.42; N, 9.48.

4.3.10. *5-(4-methoxyphenyl)indazolo[3,2-b]quinazolin-7(5H)-one* (**5j**). Yield 83% (283 mg); White solid; M.p = 220–224 °C; IR (KBr, cm$^{-1}$): 3048, 1680, 1616, 1468, 1050; $^1$H NMR (DMSO-d6, 500 MHz): δ = 3.77 (s, 3H), 6.78 (d, *J* = 7.0 Hz, 1H), 6.98 (d, *J* = 7.5 Hz, 1H), 7.15 (t, *J* = 7.5 Hz, 1H), 7.29 (t, *J* = 7.5 Hz, 1H), 7.40 (d, *J* = 7.0 Hz, 1H), 7.46 (d, *J* = 7.5 Hz, 1H), 7.53-7.77 (m, 5H-ph), 8.14 (d, *J* = 7.5 Hz, 1H); $^{13}$C NMR (DMSO-d6, 125 MHz): δ = 56.4, 111.9, 115.2, 115.6, 115.7, 116.1, 117.7, 121.6, 126.1, 128.0, 128.2, 129.0, 133.5, 140.3, 146.0, 147.1, 154.3, 163.4; Anal. Calcd for C$_{21}$H$_{15}$N$_3$O$_2$: C, 73.89; H, 4.43; N, 12.31. Found: C, 74.09; H, 4.66; N, 12.48.

4.3.11. *5-(4-bromophenyl)-4-methoxyindazolo[3,2-b]quinazolin-7(5H)-one* (**5k**). Yield 78% (328 mg); White solid; M.p = 215–219 °C; IR (KBr, cm$^{-1}$): 3068, 1674, 1625, 1468, 1068; $^1$H NMR (DMSO-d6, 500 MHz): δ = 3.82 (s, 3H), 7.01 (t, *J* = 7.5 Hz, 1H), 7.43 (d, *J* = 7.5 Hz, 2H), 7.52 (t, *J* = 8.0 Hz, 1H), 7.70 (d, *J* = 8.0 Hz, 1H), 7.72 (d, *J* = 7.5 Hz, 2H), 7.84 (d, *J* = 7.0 Hz, 1H), 7.92 (d, *J* = 7.5 Hz, 1H), 8.01 (d, *J* = 7.5 Hz, 1H), 8.30 (d, *J* = 7.0 Hz, 1H); $^{13}$C NMR (DMSO-d6, 125 MHz): δ = 56.7, 112.0, 113.8, 114.0, 115.0, 117.9, 118.9, 120.6, 126.2, 126.9, 128.0, 129.2, 131.0, 133.7, 136.7, 146.7, 154.2, 162.8; Anal. Calcd for C$_{21}$H$_{14}$BrN$_3$O$_2$: C, 60.02; H, 3.36; N, 10.00. Found: C, 6.28; H, 3.54; N, 10.18.

4.3.12. *4-methoxy-5-(4-methoxyphenyl)indazolo[3,2-b]quinazolin-7(5H)-one* (**5l**). Yield 69% (256 mg); White solid; M.p = 206-212 °C; IR (KBr, cm$^{-1}$): 3048, 1665, 1625, 1455, 1066; $^1$H NMR (DMSO-d6, 500 MHz): δ = 3.71 (s, 3H), 3.84 (s, 3H), 6.72 (t, *J* = 7.5 Hz, 1H), 7.00 (dd, *J*$_1$ = 8.5 Hz, *J*$_2$ = 3.0 Hz, 2H), 7.05 (d, *J* = 7.5 Hz, 1H), 7.16 (d, *J* = 8.0 Hz, 1H), 7.21-7.24 (m, 4H), 7.88 (d, *J* = 8.0 Hz, 1H), 8.32 (d, *J* = 7.5 Hz, 1H); $^{13}$C NMR (DMSO-d6, 125 MHz): δ = 55.3, 109.9, 113.9, 114.6, 117.4, 117.8, 122.1, 127.6, 128.2, 128.8, 132.2, 133.6, 134.6, 146.7, 149.6, 149.9, 154.2, 158.5, 162.8; Anal. Calcd for C$_{22}$H$_{17}$N$_3$O$_3$: C, 71.15; H, 4.61; N, 11.31. Found: C, 71.34; H, 4.83; N, 11.60.

**Acknowledgment**

This research was supported by a grant from Iran National Science Foundation (INSF).